\documentclass[a4paper,10pt]{IEEEtran}
\usepackage[left=1.3cm,right=1.3cm,top=1.82cm, bottom = 4.3cm]{geometry}

\usepackage{setspace}
\usepackage{multirow}
\usepackage{lipsum}
\usepackage{amsfonts}
\usepackage{array}
\usepackage{amsmath,cases,amssymb}
\usepackage{amsmath}
\usepackage{amsthm}
\usepackage{graphicx}
\usepackage{cite}
\usepackage{subfigure}
\usepackage{epsfig}
\usepackage[hyphens]{url}
\usepackage{algorithm}
\usepackage{algorithmic}
\usepackage{epstopdf}
\usepackage{color}
\usepackage{makecell}
\usepackage{xcolor}
\usepackage[normalem]{ulem}

\usepackage{mathtools}

\newcommand{\SA}{\mathtt{SA}}

\DeclareGraphicsExtensions{.eps,.pdf}

\usepackage{soul}

\makeatletter
\newcommand{\doublewidetilde}[1]{{%
		\mathpalette\double@widetilde{#1}}}
\newcommand{\double@widetilde}[2]{%
	\sbox\z@{$\m@th#1\widetilde{#2}$}%
	\ht\z@=.5\ht\z@
	\widetilde{\box\z@}}
\makeatother

\newcommand{\bPsi}{\boldsymbol{\Psi}}

\newcommand{\bepsilon}{\boldsymbol{\xi}}

\newcommand{\bw}{\mathbf{w}}

\newcommand{\bC}{\mathbf{C}}

\newcommand{\bH}{\mathbf{H}}

\newcommand{\bI}{\mathbf{I}}

\newcommand{\bu}{\mathbf{u}}

\newcommand{\bP}{\mathbf{P}}

\newcommand{\bo}{\mathbf{o}}

\newcommand{\bs}{\mathbf{s}}
\newcommand{\bK}{\mathbf{K}}

\newcommand{\bmu}{\pmb{\mu}}

\newcommand{\Qcal}{\mathcal{Q}}
\newcommand{\Kcal}{\mathcal{K}}

\newcommand{\Mcal}{\mathcal{M}}

\newcommand{\Ncal}{\mathcal{N}}

\newcommand{\Pcal}{\mathcal{P}}

\newtheorem{remark}{Remark}

\begin{document}
	

\title{Digital Twin of Industrial Networked Control System based on Value of Information}

\author{Van-Phuc Bui, \textit{IEEE Student Member}, Daniel Abode, Pedro M. de Sant Ana, Karthik Muthineni, \\ Shashi Raj Pandey, \textit{IEEE Member}, and Petar Popovski, \textit{IEEE Fellow}
		\thanks{V.-P Bui, D. Abode, S.R. Pandey, and P. Popovski (emails: \{vpb, danieloa, srp, petarp\}@es.aau.dk) are all with the Department of Electronic Systems, Aalborg University, Denmark. P. M. de Sant Ana and K. Muthineni are with the Corporate Research, Robert Bosch GmbH, 71272 Renningen, Germany (email: \{Pedro.MaiadeSantAna, karthik.muthineni\}@de.bosch.com). This work was supported in part by the Villum Investigator Grant ``WATER'' from the Velux Foundation, Denmark. The work was supported by the EU Horizon 2020 research and innovation programme: Daniel Abode and Karthik Muthineni under the Marie Skłodowska-Curie (Grant 956670) programme and Petar Popovski under the CENTRIC project (Grant 101096379).}	}

\maketitle
\begin{abstract}
The paper examines a scenario wherein sensors are deployed within an Industrial Networked Control System, aiming to construct a digital twin (DT) model for a remotely operated Autonomous Guided Vehicle (AGV). The DT model, situated on a cloud platform, estimates and predicts the system's state, subsequently formulating the optimal scheduling strategy for execution in the physical world. However, acquiring data crucial for efficient state estimation and control computation poses a significant challenge, primarily due to constraints such as limited network resources, partial observation, and the necessity to maintain a certain confidence level for DT estimation. We propose an algorithm based on Value of Information (VoI), seamlessly integrated with the Extended Kalman Filter to deliver a polynomial-time solution, selecting the most informative subset of sensing agents for data. Additionally, we put forth an alternative solution leveraging a Graph Neural Network to precisely ascertain the AGV's position with a remarkable accuracy of up to 5 cm. Our experimental validation in an industrial robotic laboratory environment yields promising results, underscoring the potential of high-accuracy DT models in practice. 
\end{abstract}

\vspace*{-0.5cm}
\section{Introduction}\label{sec:intro}
\vspace*{-0.1cm}

\begin{figure}[t!]
	\centering
	\includegraphics[width=0.47\textwidth]{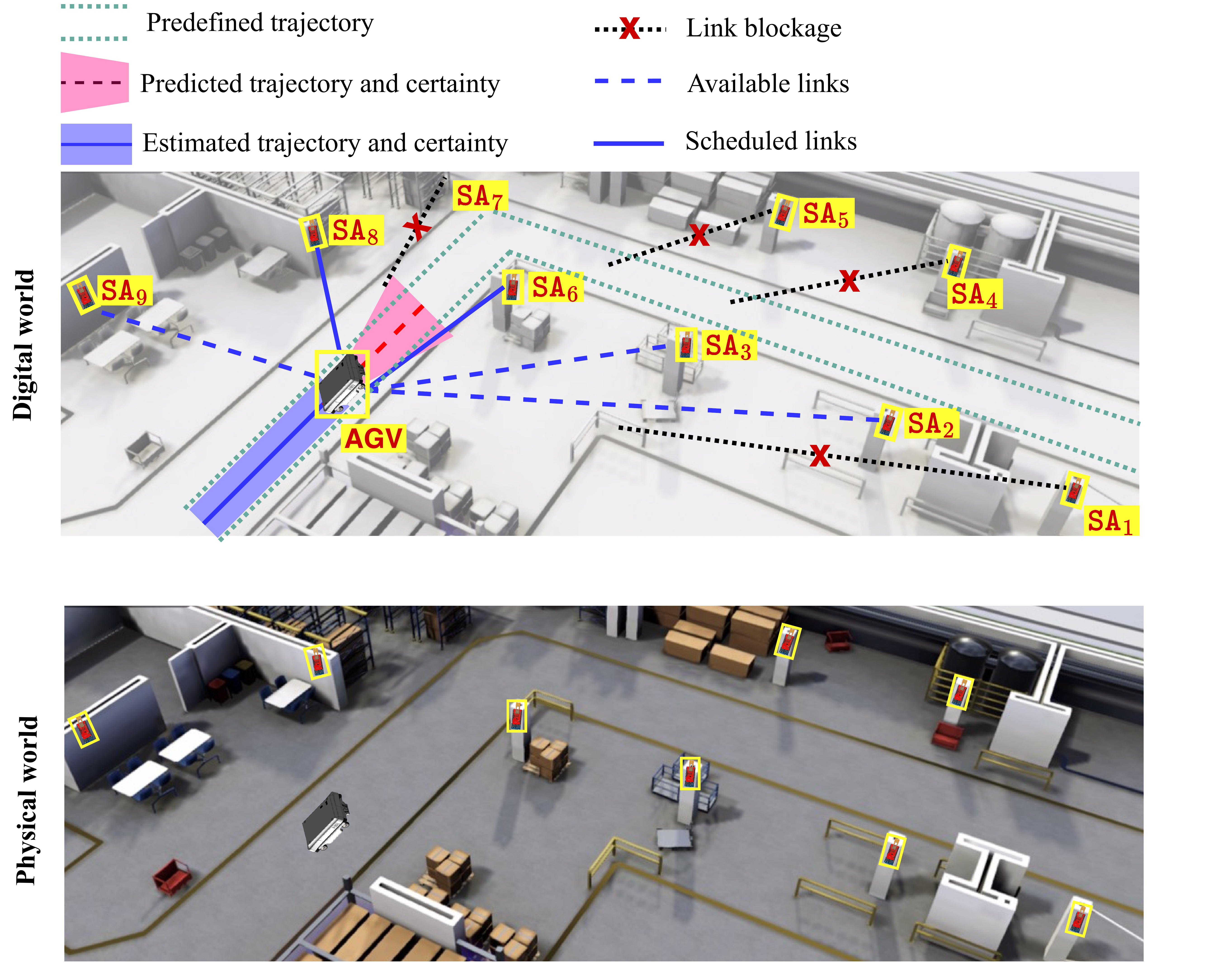}
 \vspace*{-0.3cm}
	\caption{The considered architecture with a Digital Twin (DT).}
	\label{fig_system_model}
	\vspace*{-0.6cm}
\end{figure}

In Industry 4.0 smart manufacturing, vast amounts of real-time data from wireless sensors are essential \cite{tang2015tracking}. Digital twin (DT) models, unlike traditional simulators or optimization tools, convert this data into predictive models, aiding real-time decision-making~\cite{9899718}. A promising use case, an industrial Network Control System (NCS) representing the physical world can integrate DTs in the edge or at the cloud by collecting data from sensor devices and other central/distributed, allowing real-time operational decisions. For instance, consider localization and tracking problems in NCS with DT and the challenges therein to control the maneuver of the moving object. Fig.~\ref{fig_system_model} depicts a dynamic process system where DTs estimate the state of an Autonomous Guided Vehicle (AGV). \emph{Sensing agents}  ($\SA$s) transmit noisy and partial observations to the Cloud-connected AGV for DT model development \cite{li2020digital}. After virtual processing, i.e., the model is updated, the next system state is predicted, and the optimal policy is computed. However, predictions guide optimal policy decisions, which eventually have an impact on physical operations. This hints at requirements for timely collecting  data to make accurate predictions with the DT model and, therein, precisely control actuation in the physical world.


A compelling connectivity solution for sensory data collection to build DT models is Ultra-Wideband (UWB) technology. Amongst all types of RF-based system tracking technologies based on the various standards (IEEE 802.11, IEEE 802.15.4, RFID, Bluetooth, GSM, WiFi), UWB provides energy savings and high accuracy efficiency due to its good time domain resolution \cite{oppermann2004uwb, dwek2019improving, guo2019ultra}. However, it is essential to acknowledge the sensitivity of UWB to Line-of-Sight (LoS) links, where Non-Line-of-Sight (NLoS) scenarios may compromise accuracy. Besides connectivity issues in practical scenarios, to enhance the precision of state monitoring in AGVs, Inertial Measurement Units (IMUs) are often used. IMUs are equipped with sensors such as accelerometers and gyroscopes and could offer precise orientation and acceleration measurements, thus supplementing UWB's data for the DT model \cite{ahmad2013reviews}. There are several works on using UWB and/or IMUs for state monitoring \cite{dwek2019improving, guo2019ultra, sczyslo2008hybrid}. While studies \cite{guo2019ultra} and \cite{sczyslo2008hybrid} emphasized the efficient utilization of UWB and IMU signals to enhance localization, the authors in \cite{dwek2019improving} devised techniques to identify and eliminate outliers for more accurate localization. Furthermore, recent advancements include the application of novel machine learning approaches such as Deep Neural Networks (DNN) \cite{10054386} and Graph Neural Networks (GNN) \cite{Justin2017} to address localization challenges. 
However, the related works fail to assess the significance and relevance of each feature within AGV's state space for accurate state estimation. Furthermore, several factors in the industrial environment, such as when the number of UWB anchors is extensive or sparsely distributed, add unique challenges for data collection and its use for localization. In most cases, many of these anchors provide non-LoS information, resulting in very high error measurements that could potentially lead to inaccuracies in object tracking. In addition, UWB receivers are engineered to handle a limited number of UWB signals concurrently, underscoring the significance of selecting the appropriate anchors to maximize the efficient utilization of wireless resources. Therefore, jointly using UWB and IMU signals within the complex environment of industrial NCS presents fundamental challenges of potentially high computational time, inefficient use of available communication resources, and significant measurement discrepancies that eventually lead to errors in object tracking with the DT solution. 

In this paper, we propose a novel DT framework to streamline the selection and arrangement of UWB anchors for AGV state monitoring with confidence guarantees on the estimation. This approach aims to strategically schedule anchors, particularly in scenarios with redundant or obstructed anchors that are unnecessary for state monitoring. 
Our contributions are listed as follows: \textit{(i)} We develop a DT framework applicable to industrial manufacturing environments for monitoring the state of AGVs and formulate a problem to select suitable active $\SA$s to maintain the confidence of the system estimate of the DT; \textit{(ii)} We propose a Value of Information (VoI)-based algorithm for efficient and practical implementations of decision-making under uncertainty, yielding a low-complexity solution in polynomial time; \textit{(iii)} In addition, we developed a GNN-based sub-optimal localization solution that addresses scalability and complexity issues; and to that end, \textit{(iv)} We validate the results through real-world experiments conducted in the industrial laboratory.

\begin{figure}[t!]
    \centering
    \includegraphics[width=0.4\textwidth]{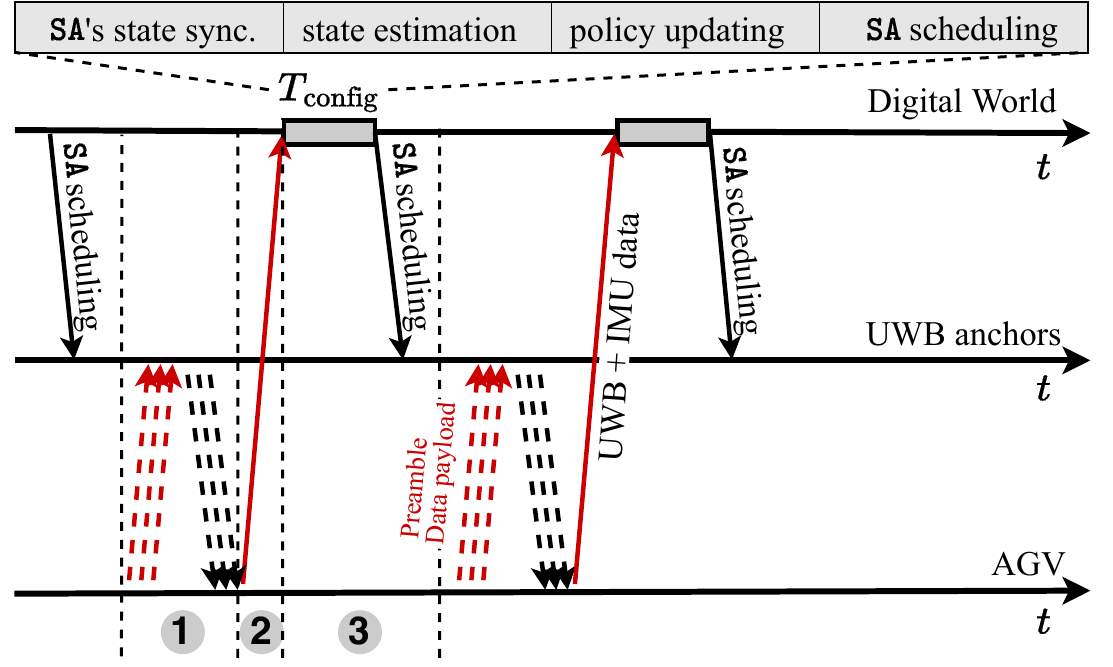}
        \vspace*{-0.2cm}
    \caption{The system's communication diagram, includes 3 main phases: (1) sensing; (2) data uploading; and (3) policy downloading and distributing.}
    \label{fig_tdma}
    \vspace*{-0.6cm}
\end{figure}
\vspace*{-0.26cm}
\section{Digital Twin Architecture}
\vspace*{-0.1cm}
We consider a WSN, as in Fig.~\ref{fig_system_model}, including one AGV integrated with a UWB tag, one IMU sensor and a set of $\Mcal=\{1,2,\dots, M\}$ UWB anchors. These anchors are denoted as \textit{sensing agents} $(\SA)$s, which are synchronized and controlled by the DT through a time-slotted wireless channel for building the DT model of AGV, where each \emph{query interval} (QI) occurs at $n \in \Ncal= \{1, 2, \dots, N\}$. The UWB tag-anchor pairs provide observations to ascertain AGVs' state, featuring an active-inactive mechanism \cite{oppermann2004uwb}. This feature holds significance in expansive indoor environments housing numerous UWB tags dedicated to precise positioning tasks. In scenarios where only a few AGVs operate within a confined space, selectively activating UWB anchors serves a dual purpose: conserving energy and avoiding interference with concurrent operations. Based on the current model state and the updated information on the environment, the DT model predicts the AGV's state and compute the optimal policy on scheduling anchors. The feedback is sent to prompt actions within the physical world. In this paper, uppercase and lowercase bold letters are matrices and vectors, respectively. $\mathbf{x}\sim\mathcal{CN}(\boldsymbol{\mu}, \boldsymbol{\Sigma})$ is a random vector $\mathbf{x}$ conforming to a complex circularly symmetric Gaussian distribution with a mean of $\boldsymbol{\mu}$ and a covariance matrix of $\boldsymbol{\Sigma}$. 

\begin{figure}[t!]
	\centering
	\includegraphics[width=0.35\textwidth]{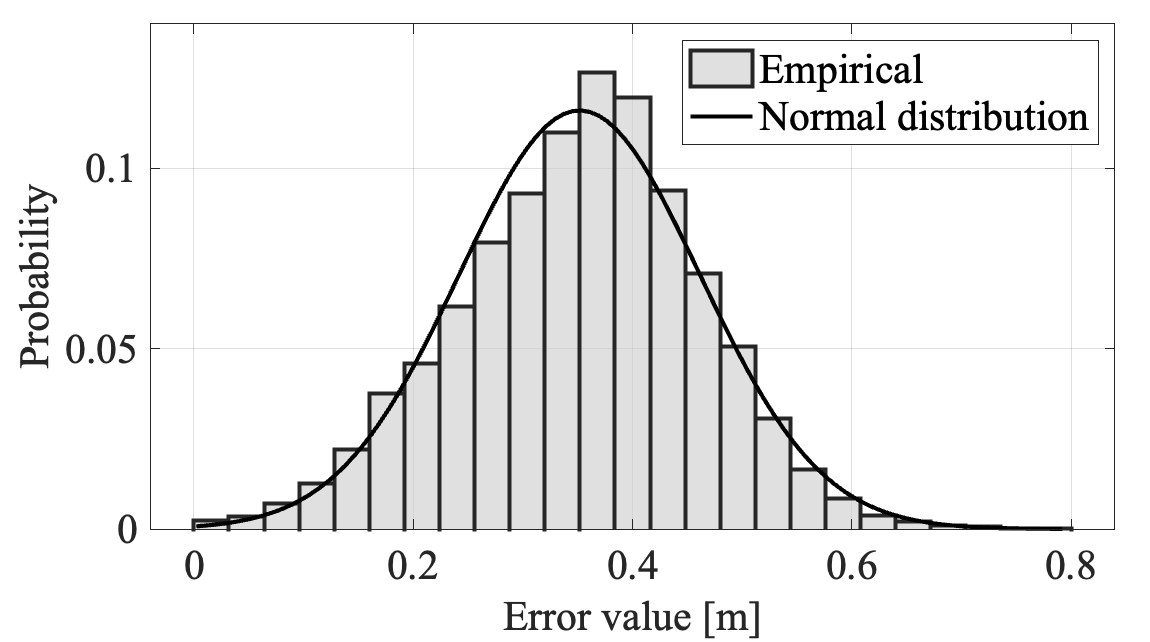}
        \vspace*{-0.3cm}
	\caption{Experimental errors distribution and theoretical approximation.}
	\label{fig_UWB_error}
	\vspace*{-0.65cm}
\end{figure}

The AGV functions within a $K$-dimensional process denoted as $\mathcal{K}={1,2,\dots,K}$, where its state at time $n$, denoted as $\mathbf{s}(n) = [s_1(n), s_2(n), \dots, s_K(n)]^\top$, evolves by
\begin{align}\label{state_model}
	\bs(n) &= f(\bs(n-1)) + \bu(n), \forall n\in\Ncal, 
\end{align}
where $f:\mathbb{R}^{K}\rightarrow\mathbb{R}^K$ is the state update function and $ \bu(n)\sim \mathcal{N}(\mathbf{0},\mathbf{C}_{\bu})$ stands for the process noise. 
At QI $n$, a $D$-dimensional observation  provided by  $\SA_m$  $(m \in \mathcal{M})$ is 
\begin{align}\label{obs_model}
	\bo_{m}(n) = g(\bs(n))+ \mathbf{w}_{m},
\end{align}
where $\bw_m\sim \Ncal(\boldsymbol{\mu}_{\bw_m}, \mathbf{C}_{\bw_m})$ stands for the measurement noise. Generally, the covariance matrices $\mathbf{C}_{\bu}$ and $\mathbf{C}_{\bw_m}$ are non-diagonal. Additionally, we assume that the sets $\{\{\bs(n)\}_{\forall n},\{\bu(n)\}_{\forall n}, \{\bw_m(n)\}_{\forall m, n}\}$ are uncorrelated, and that $f$, $\mathbf{C}_{\bu}$, and  $\mathbf{C}_{\bw_m} \forall m$ are known.

\vspace*{-0.3cm}
\subsection{Communication System}
\vspace*{-0.1cm}

The communication between the AGV and DT uses a Time Division Multiple Access (TDMA) access scheme, with alternating downlink and uplink phases as illustrated in Fig.~\ref{fig_tdma}. Each QI comprises three primary phases, commencing with the sensing phase, during which scheduled $\SA$s observe the environment and transmit their observations to the AGV; these observations are subsequently relayed to the DT via a wireless link in phase two. At the DT, $T_\text{config}$ encompasses the time required for the DT model to update the active state of $\SA$s. Subsequently, it estimates the full state and schedules a maximum of $C$ $\text{SA}$s during the downlink phase. 

\textit{UWB sensors:} Time of Arrival (ToA) system are used to estimate the distance between the AGV and $\SA_m$ at QI $n$, where the moving AGV equipped UWB tag and UWB anchor $m$ are positioned at $P_\text{AGV} = [x, y, z]^\top$, and $P_m = [x_m, y_m, z_m]^\top $, respectively. The distance in 3D space between the AGV and $\SA_m$ is denoted as $d_{m} = ||P_\text{AGV} - P_m||_2$. When combined with the observation model described in \eqref{obs_model}, the observation from a UWB anchor $m$ can be represented as $o_{m}^\mathtt{uwb}(n) = d_m(n) + w_m^\mathtt{uwb}$. In this expression, $w_m$ represents the noise inherent in our experimental setup, as depicted in Fig.~\ref{fig_UWB_error}, which closely resembles a Gaussian distribution. Further specific experimental setups are detailed in Section~\ref{sec:experiment}.

\textit{IMU sensor:} A typical IMU sensor is equipped with a 3-axis accelerometer and a 3-axis gyroscope, which provide acceleration and angular velocity measurements, respectively. Following processing steps, the IMU signal could be modeled as $\bo_m^\mathtt{imu}(n) = \mathbf{a}(n) + \bw^\mathtt{imu}_m$, where $\bw^\mathtt{imu}_m \sim \mathcal{N}(\mathbf{0}, \mathbf{C}_{\bw_m^\mathtt{imu}})$, and $\mathbf{a}(n)$ indicating accelerations from $x$ and $y$ axes. The recorded accelerations correspond to the local coordinate frame of the IMU sensor. To integrate the IMU data in the AGV state estimation process, the accelerations from the IMU sensor should convey the movement of the AGV in a global frame of reference to which the other sensor data on the AGV are being recorded and monitored. Using the yaw angle $\psi$, the accelerations in the local coordinate frame $\mathbf{a}_l(n)$ can be mapped to the global coordinate frame $\mathbf{a}_g(n)$ as
\begin{equation}
\begin{bmatrix}
a_{x} \\
a_{y} 
\end{bmatrix}_{g}
=
\begin{bmatrix}
\cos({\psi}) & \sin({\psi}) \\
-\sin({\psi}) & \cos({\psi})  
\end{bmatrix}
\begin{bmatrix}
a_{x} \\
a_{y} 
\end{bmatrix}_{l}.
\end{equation}

\vspace{-15pt}
\subsection{Problem Formulation}
The goal of the DT model is to ensure a precise assessment of the AGV's state over its belief of states. In this paper, the estimator $\hat{\bs}(n)$ for ${\bs}(n)$ is modeled as $p(\bs(n))\sim \mathcal{N}(\hat{\bs}(n), \bPsi(n)), n \in\Ncal$, where the covariance matrix $\bPsi(n)$ is adjusted at QI $n$ through an Extended Kalman Filter (EKF).
The MSE of the estimator is
\begin{equation}\label{}
	\text{MSE}_{} = \mathbb{E}\big[||\bs(n)-\hat{\bs}(n)||^2_2\big], n \in\Ncal. 
\end{equation}
\begin{remark}\label{certainty}
	{\color{black}
		We define the maximum acceptable standard deviation for feature $k\in\Kcal$ as $\xi_k$. This corresponds to the following condition:
		\begin{equation}\label{qos_condition}
			[\bPsi(n)]_{k} \leq {\xi}^2_{k}, \forall k \in\Kcal,
		\end{equation}
		where $[\bPsi(n)]_{k}$ is the $k$-th element of the diagonal of $\bPsi(n)$.}
\end{remark}

\begin{algorithm}[t]
	\begin{algorithmic}[1]{\fontsize{8pt}{9pt}\selectfont
			\protect\caption{$\SA$ scheduling algorithm for problem  \eqref{glob_problem}} 
		\label{alg_global}
		\global\long\def\algorithmicrequire{\textbf{Input:}}
		\REQUIRE $C_{\bu},\bmu_{\bw_m}, C_{\bw_m},$
		Available uplink slots $C$, The state and requirement certainty $\big(\bs, \{\xi_{k}^2\}\big)$
		\global\long\def\algorithmicrequire{\textbf{Output:}}
		\REQUIRE The scheduled user set $\{\Qcal^*(n)\}$; their belief $\{\hat{\bs}^*(n), \bPsi^*(n)\}$
		\STATE Initial $\mathcal{Q}(n) = \emptyset$
		\STATE Compute the prior errors $\bPsi^{\mathtt{pr}}(n)$ as in \eqref{prior_error} 
		\IF {$[\bPsi^{\mathtt{pr}}(n)]_{k} \leq \xi_{k}^2, \forall k$}
		\STATE Compute ${\hat{\bs}^{\mathtt{pr}}(n)} $ as in \eqref{mu_prior}
		\STATE Update ${\hat{\bs}(n)}  = {\hat{\bs}^{\mathtt{pr}}(n)} $ and $\bPsi_{\hat{\bs}}(n) = \bPsi_{\hat{\bs}}^\mathtt{pr}(n)$
		\ELSE
		\STATE Set $t=1$ and compute available $\SA$ set $\Pcal(n)$ with DT model
		\WHILE{conditions \eqref{checking3} hold}
		\STATE Update $\Qcal(n)$ and $\Pcal(n)$ as in \eqref{update_set}
		\STATE Update the $\bK(n)$, $\bH(n)$ and $C_{\bw(n)}$ as in \eqref{H_compute}, and \eqref{Cw_compute}
		\STATE  Set $t=t+1$
		\ENDWHILE
		\STATE Update $\Qcal^*(n) = \Qcal(n)$
		\STATE Compute ${\hat{\bs}(n)}  = {\hat{\bs}^{\mathtt{pos}}(n)} $,  $\bPsi_{\hat{\bs}}(n) = \bPsi_{\hat{\bs}}^\mathtt{pos}(n)$ as in\eqref{pos_update},   \eqref{pos_variance}
		\ENDIF
	}
\end{algorithmic}
\end{algorithm}

The AGV faces operational challenges stemming from sensor placement when UWV anchors situated in critical locations, such as those obscured by obstacles or positioned at a considerable distance, may remain inactive, leading to compromised sensing accuracy. There are efforts dedicated to mitigating the inaccuracies stemming from NLoS signals emitted by UWB sensors \cite{dwek2019improving}. However, UWB requires LoS as any commercially available positioning system. When obstructed, expect inaccuracies similar to the size of the obstacle, if not greater. In some cases, tracking may be severely compromised or completely incorrect when LoS is obstructed, leading to potential fluctuations in sensing reliability. To ensure the seamless and dependable operation of AGVs within industrial settings, we use DT to predict and activate UWB anchors capable of establishing LoS connections. By harnessing predictive analytics, we can anticipate the spatial distribution of obstacles and optimize the deployment of UWB anchors accordingly, thereby enhancing operational predictability and reliability.

At the DT models, we delineate the available subset of $\SA$s at QI $n$ as $\mathcal{P}(n)$. The scheduling decision hinges on the anticipated expected VoI associated with each $\SA$, crucial for refining the AGV state estimates \cite{box2011bayesian}. This frames a scheduling problem for the $\SA$s, aiming to select those that actively contribute sensing signals while fulfilling certainty constraints \eqref{qos_condition}. We introduce the scheduling set $\mathcal{Q}(n)=\{\SA_m\in\mathcal{P}(n)\}$, where $\mathcal{P}(n)$ denotes the available $\SA$ set at QI $n$. To ensure compatibility within a QI, we establish a maximum number of connections, $C$, aligning with the uplink slots available before the subsequent QI. The scheduling problem is then defined as
\begin{subequations} \label{glob_problem}
	\begin{alignat}{2}
		\Qcal(n)^*=&\  \underset{\Qcal(n)}  {\arg \min}& &\sum_{k\in\Kcal}\max\left\{\frac{[\bPsi(n)]_{k }}{\xi^2_{k}}-1,0\right\} \label{glob_problema} \\
		&\mbox{subject to } && |\Qcal(n)| \leq C, \label{glob_problemb}\\
            &&& |\Qcal(n)| \subseteq \Pcal(n). \label{glob_problemc}
	\end{alignat}
\end{subequations}
The objective function \eqref{glob_problema} aims to minimize the MSE of the DT model, while the constraint \eqref{glob_problemb} ensures that the number of active $\SA$s does not exceed the maximum number of connections $C$.  It is clear that by querying observations from more $\SA$s, the estimation accuracy increases. For those $\SA$s with significant errors in their measurements, i.e., those blocked by obstacles or too far away, measuring and sending observations is inefficient. 
The issue \eqref{glob_problem} arises from the non-convex nature of both the objective function \eqref{glob_problema} and the constraints \eqref{glob_problemb} and \eqref{glob_problemc}. Additionally, the node selection aspect renders the problem analogous to the well-known NP-hard knapsack problem. Consequently, a heuristic algorithm is employed to derive a practical yet suboptimal solution.

\begin{figure*}[t]
	\centering
	\includegraphics[trim=0cm 2.2cm 1cm 0.2cm, clip=true, width=0.9\textwidth]{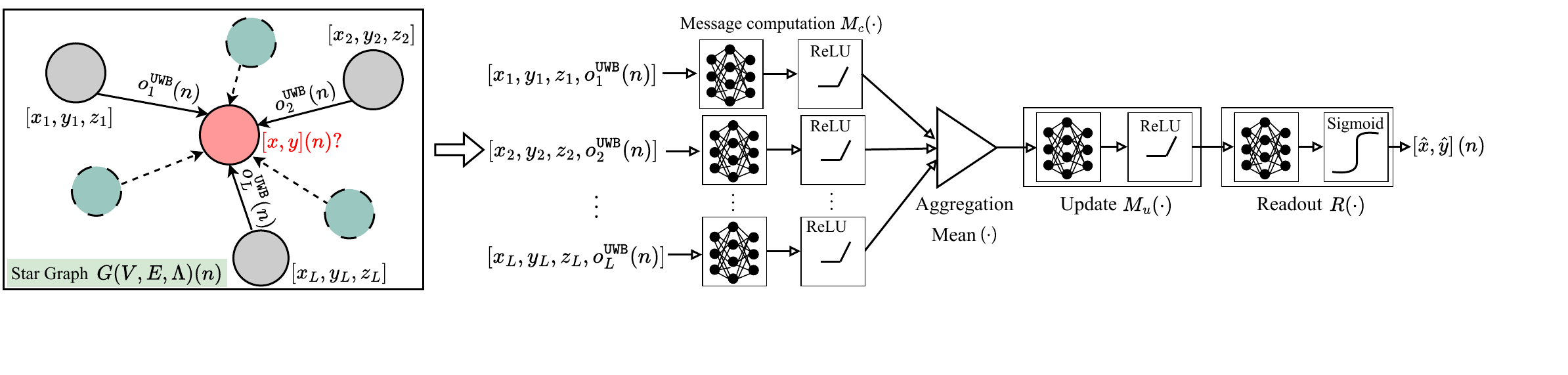}
        \vspace*{-0.3cm}
	\caption{GNN architecture for AGV positioning.}
	\label{fig_GNN}
	\vspace*{-0.6cm}
\end{figure*} 

\vspace{-5pt}
\section{VoI-based $\SA$ Scheduling}
The EKF is prevalently utilized in IoT literature~\cite{bui2023value}. Additionally, it is often assumed that the virtual environment possesses comprehensive knowledge regarding the statistical attributes of the process, including the update function $f(\mathbf{s})$ and noise covariance matrices. This assumption is commonly found in the relevant literature, as system statistics estimation can typically be conducted prior to deployment \cite{8648169}. The fundamental approach to addressing \eqref{glob_problem} involves selecting the minimum number of $\SA$s at QI $n$ to transmit data, ensuring the desired level of state estimation certainty for state $\bs(n)$. This is achieved by identifying the subset of $\SA$s $\mathcal{Q}^*(n) \subseteq \Pcal(n)$ with the smallest cardinality that satisfies constraint \eqref{glob_problemb}, while ensuring that all selected features meet the condition outlined in equation \eqref{qos_condition}. The heuristic algorithm, outlined in pseudocode in Algorithm \ref{alg_global}, effectively resolves problem \eqref{glob_problem}. 
The initial state $\bs(n)$ represents a random vector with a defined mean $\mathbb{E}[\bs(n)] = \boldsymbol{\mu}_{\bs(0)}$ and covariance matrix $\text{Cov}[\bs_0]= \mathbf{C}_{\bs(0)}$. At the outset, $\Qcal(n)$ is initialized as empty due to the absence of prior information. EKF computes the estimation errors for the belief $\hat{\bs}_t$, following a normal distribution $\mathcal{CN}(\bmu_{\hat{\bs}_t},\bPsi^{\mathtt{pr}}_t)$ at the AGV, based on preceding updates $\hat{\bs}_{t-1}$ as 
\begin{align}\label{prior_error}
    \bPsi^{\mathtt{pr}}_{\hat{\bs}_t} = \bP\bPsi_{\hat{\bs}_{t-1}}\bP^\top + \bC_{\bu_t},
\end{align}
where the Jacobian matrix $\bP= \mathcal{J}\{f(\bs({n-1}))\}$  linearizes the nonlinear model of $f(\bs({n-1}))$.
For a given set of error variance qualities $\bepsilon \triangleq \{\xi_{k}\}_{k\in\Kcal}$, the conditions outlined in \eqref{qos_condition} lead to two possible scenarios: \textit{(i)} If conditions \eqref{qos_condition} are satisfied for all $k\in\Kcal$, it indicates that the DT model meets the required bound without necessitating any observations from $\SA$s. Then, the prior update alone ensures the confidence in the estimate, and thus $\Qcal^*(n) = \emptyset$. \textit{(ii)} Conversely, if any of these conditions are not met, at least one feature is inaccurately estimated, warranting the $\SA$ observations to enhance the estimation process, as scheduled by our heuristic. The belief in the first case is computed using the EKF blind update operation as
\begin{equation}\label{mu_prior}
	{\hat{\bs}^{\mathtt{pr}}(n)}  = \bP{\hat{\bs}(n-1)} + \bmu_{\bu}.
	\vspace{-0.1cm}
\end{equation}
In the second case, we execute Alg.~1. We tress that if any constraint remains unfulfilled and $\{|\Qcal(n)| < C, |\mathcal{P}(n)| >0\}$ at the $t$-th iteration, there is chance to schedule new sensing agents $\SA$s to join $\Qcal(n)$. To do it, we select the predicted closed $\SA_m\in\Pcal(n)$ at iteration $t$ as
\begin{equation}\label{finding_state}
    \SA_m^* =  \underset{\SA_m \in \Pcal(n)}{\arg \min}d_m(n) \triangleq ||\hat{P}_\text{AGV}(n)-P_m||_2,
\end{equation}
where $\hat{P}_\text{AGV}(n)$ is AGV's estimated position QI $n$.
The scheduled and available $\SA$ sets $\Qcal(n)$ and $\mathcal{P}(n)$ are updated by
\begin{equation}\label{update_set}
    \Qcal_t \leftarrow \Qcal_t \cup\{ \SA_m^* \};\  \Pcal_t \leftarrow\Pcal_t\backslash\{\SA_{{m}}^*\}.
\end{equation}
$\bH(n)$ and $\bC_{\bw(n)}$ are the combination observation and covariance matrices, which are respectively formulated as
\begin{align}
\bH(n) &= [\bH_{1};\bH_2;\dots;\bH_{|\Qcal(n)|}], \label{H_compute} \\
\bC_{\bw(n)} &=  \text{diag}[\bC_{\bw_1}, \bC_{\bw_2},\dots,\bC_{\bw_{|\Qcal(n)|}}]\label{Cw_compute},
\end{align}
where $\bH_{{m}}= \mathcal{J}\{g(\bs({n}))\}$ is the Jacobian matrix linearizing the nonlinear observation function of $g(\bs({n}))$. In this setup, the observation vector could be formulated from $o_{m}^\mathtt{uwb}(n)$ as $\mathbf{H}_m(n) = [\frac{\hat{x}-x_m}{o_{m}^\mathtt{uwb}(n)}, 0, \frac{\hat{y}-y_m}{o_{m}^\mathtt{uwb}(n)}, 0]^T$. Then, the posterior error covariance matrix is 
\begin{align}\label{pos_variance}
\bPsi^{\mathtt{pos}}_{\hat{\bs}}(n)=  (\bI - \bK(n)\bH(n))\bPsi^{}_{\hat{\bs}}(n-1),
\end{align}
where $\bK(n) =  \bPsi^{\mathtt{pr}}_{\hat{\bs}(n)}{\bH}(n)^T\big(\bC_{\bw(n)}  + {\bH}(n) \bPsi^{\mathtt{pr}}_{\hat{\bs}(n)}{\bH}(n)^T\big)^{-1}$ is the EKF gain, computed using the EKF equation. We run the iterative loop continues while all the following conditions hold:
\begin{align}
\{|\Qcal^*(n)| &< C;\  
\exists [\bPsi(n)]_{k} \geq {\xi}^2_{k}; \ 
\exists s_k^{*}(n) \mbox{ in }\eqref{finding_state}\}. \label{checking3}
\end{align}
It becomes evident that the loop iterates for a maximum of $C$ iterations before reaching termination. Subsequently, the posterior update is computed as
\begin{equation}\label{pos_update}
{\hat{\bs}^{\mathtt{pos}}(n)}  = {\hat{\bs}^{\mathtt{pr}}(n)} + \bK(n)(\bo(n) - \bH(n){\hat{\bs}^{\mathtt{pr}}(n)} ),
\end{equation}
where $\bo(n) $ represents the combination of received $\SA$ observations. Consequently, we update $\hat{\bs}(n) = \hat{\bs}^{\mathtt{pos}}(n)$. Our methodology guarantees a sustainable equilibrium between state certainty and communication expenses, tailored to the AGV's specifications, despite the diverse local scheduling solutions offered by $\SA$ for various QIs.

\vspace*{-0.2cm}
\section{DT-based GNN for AGV Localization}
\vspace*{-0.05cm}

Despite the effective solution proposed in Alg.~1, it necessitates updating the scheduling policy iteratively until a fixed point solution is reached. Additionally, applying the EKF to achieve high accuracy demands a profound understanding of the system dynamics, which can be burdensome in certain practical scenarios. In this subsection, we introduce a GNN model capable of predicting AGV positions with extremely low computational complexity and significant potential for scalability. Unlike traditional DNN/CNN techniques, GNNs handle variable-sized graph inputs without the need for preprocessing. This capability is crucial in the localization problem under practical industrial considerations, as illustrated in Fig.~\ref{fig_system_model}, where ensuring all UWB anchors are active at all times is impractical and unnecessary due to sensing range limitations and obstacles.

A forward pass in GNN consists of the message-passing phase and the readout phase. The authors in \cite{Justin2017} proposed the commonly adopted sequential framework for the message-passing phase, which includes the message computation, message aggregation, and update function. Each message-passing phase describes a GNN layer. At the end of the message-passing, each node and/or edge in the graph takes on a new representation. The readout phase processes the new representation to suit the required task, which could be a graph-level task, a node-level task or an edge-level task.  We represent the $\SA$s and AGV as nodes in a star graph, $G(V, E,\Lambda)$ of $L+1$ nodes, where the scheduled set of $L \leq M$ $\SA$s at QI $n$, $\SA_m \in \mathcal{Q}(n)$ are nodes with outgoing edges to a single node that represents the AGV. These nodes are represented with continuous circles in Fig.~\ref{fig_GNN} while the unscheduled $\SA_m \notin \mathcal{Q}(n)$ are represented with dashed circles. We index the set of $\SA_m \in \mathcal{Q}(n)$ with $l \in \{1,2, \ldots L\}$.
Let $V = \{v_1, v_2, \ldots, v_{L+1}\}$ be the set of scheduled nodes, where $v_{L+1}$ represents the AGV and $v_l \in \{v_1, v_2, \ldots, v_L\}$ represent the $\SA$s. The set of edges $E$ consists of $L$ edges, each connecting a scheduled $\SA$ node to the AGV node. The edge set $E$ can be represented as: $E = \{(v_l, v_{L+1}) \, | \, l = 1, 2, \ldots, L\}$, where each edge $(v_l, v_{L+1})$ represents a directed edge from node $v_l$ to the node $v_{L+1}$. $\Gamma \in \mathbb{R}^{(L+1) \times (L+1)}$ is a matrix of graph attributes, where $\Gamma_{l,l} = P_l, \ l = 1, 2, \ldots, L$. $\Gamma_{L+1,L+1}$ is a dummy value to represent the unknown $P_{AGV}$. $\Gamma_{l,L+1} = o_l^\mathtt{UWB}(n)$ is the measured distance from the anchor $l$ to the AGV. 

\begin{figure}[t!]
	\centering
	\includegraphics[width=0.45\textwidth]{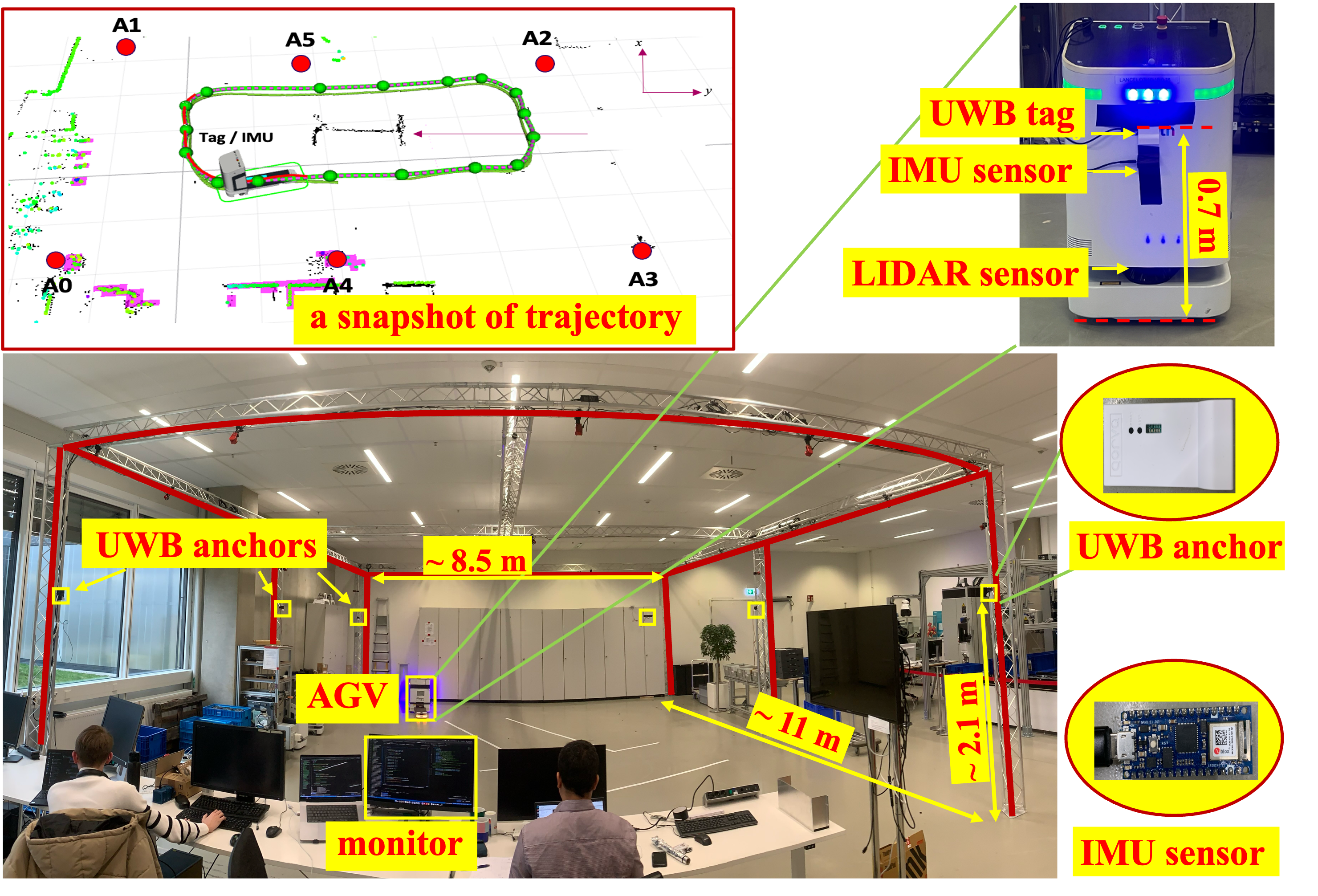}
        \vspace*{-0.3cm}
	\caption{The snapshot of the experimental environment.}
	\label{fig_ENV}
	\vspace*{-0.5cm}
\end{figure}

The forward pass architecture of our proposed GNN solution is shown in Fig.~\ref{fig_GNN}. Given the input star graph with a diameter of $2$, we consider a single-layer GNN where the node of interest is the AGV node $v_{L+1}$. The message computation $M_c(\cdot)$ is done by transforming the concatenation of $[x_l, y_l, d_{l,L+1}]$ using a multi-layer perceptron (MLP) and a rectified linear unit (ReLU). The transformed messages are aggregated at node $v_{L+1}$ and then transformed during the update phase using another MLP + ReLU layer ($M_u(\cdot)$). Finally, in the readout phase, a normalized prediction of $[\hat{x}, \hat{y}]$ is done for the corresponding QI $n$ using a Sigmoid activation function. 

We consider a supervised learning approach to train the GNN, where the set of trainable parameters is $\Pi$. The model's trainable parameters are updated using mini-batch gradient descent, with batches randomly sampled from the training dataset. After a forward pass of a batch of $\mathcal{D}$ graphs, the loss $L(\Pi)$ is computed using the expectation of the square error between the estimate $[\hat{x}, \hat{y}](d)$ and the ground truth $[x, y](d)$ $\forall \ d \in\mathcal{D}$ as in $L(\Pi) = \mathbb{E}_{\mathcal{D}}\big[\big([x,y](d)-[\hat{x}, \hat{y}](d)\big)^2\big]$.
Then, $\Pi$ is updated for the next iteration $t+1$ given a step size of $\varepsilon$ as in  $\Pi^{t+1} = \Pi^{t} - \varepsilon \nabla_{\Pi^t} L(\Pi^t),$ 
where $\nabla_{\Pi^t} L(\Pi^t)$ is the gradient of $L(\Pi)$ with respect to $\Pi$ at time $t$. The training terminates when $\Pi^{t+1} - \Pi^{t} < \epsilon$ with $\epsilon$ as an error tolerance.

\begin{figure}[t]
	\centering
	\begin{minipage}{0.49\columnwidth}
		\centering
		\includegraphics[trim=3.5cm 1cm 1cm 1cm, clip=true, width=1\textwidth]{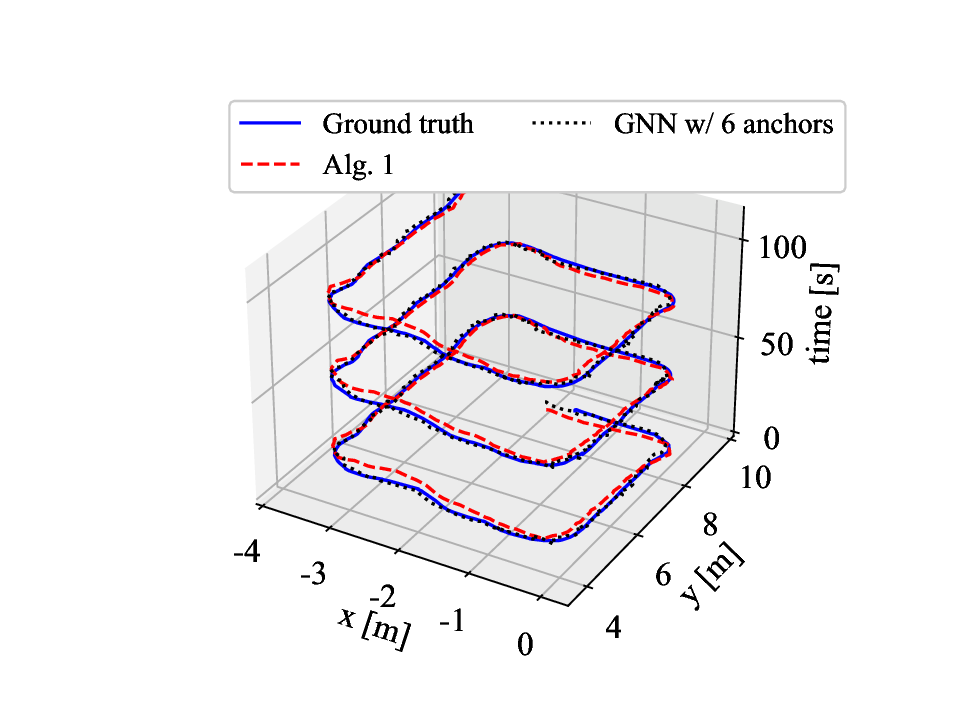} 
            \vspace{-0.5cm}
            \caption{The AGV's trajectory monitoring by Alg.1 and GNN. }
            \label{fig_trajectory}
	\end{minipage}
	\begin{minipage}{0.49\columnwidth}
		\centering
		\includegraphics[trim=0cm 0cm 0cm 0cm, clip=true, width=1\textwidth]{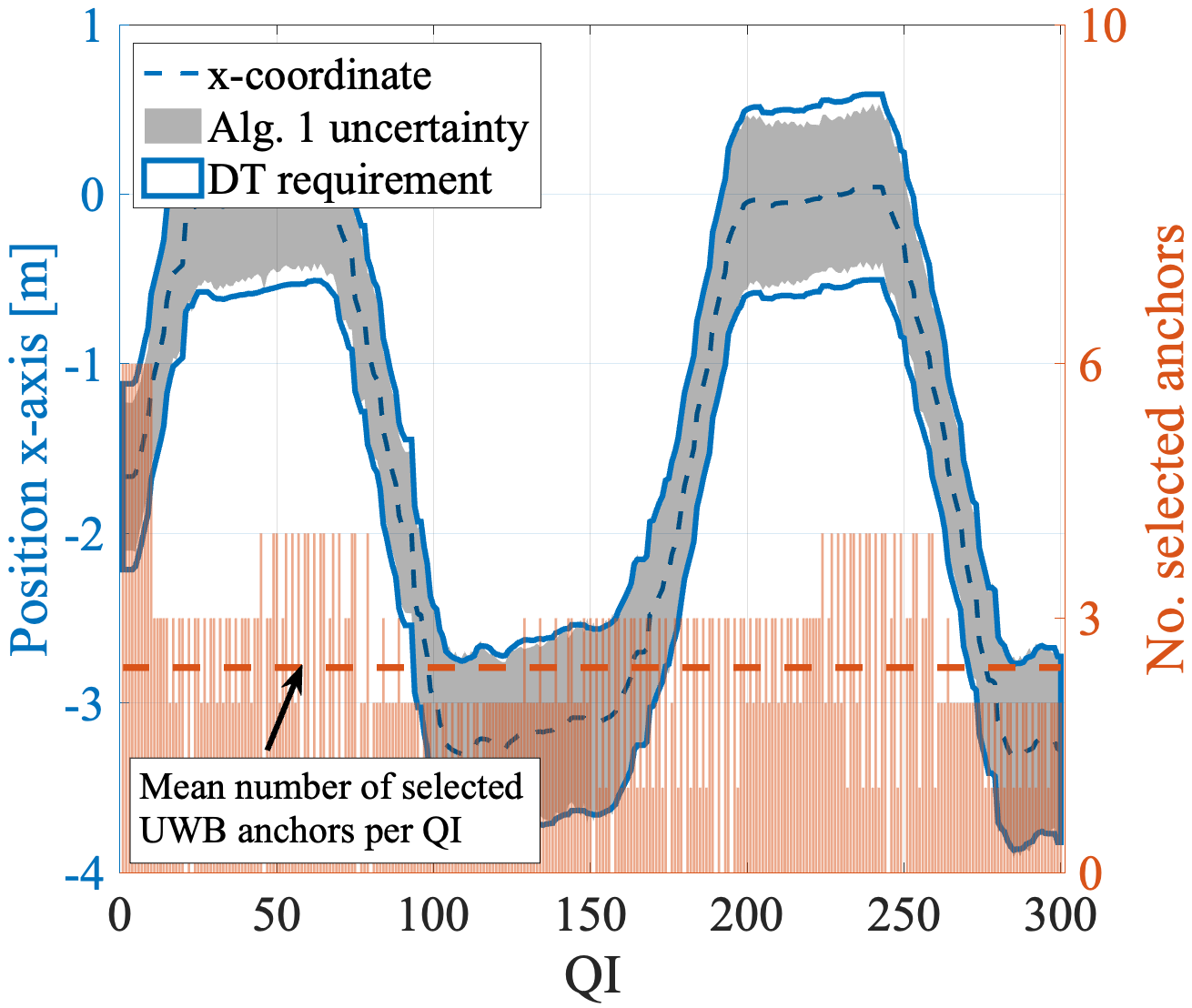} 
            \vspace{-0.5cm}
            \caption{Snapshot of uncertainty evolution and no. selected $\SA$s vs QI.}
            \label{fig_uncertainty_vs_QI}
	\end{minipage}
	\vspace{-0.5cm}
\end{figure}

\vspace*{-10pt }
\section{Experimental Evaluation}\label{sec:experiment}

\textbf{Experimental Setup.} The experimental setup consists of a platform with one example of AGV trajectory as illustrated in Fig.~\ref{fig_ENV}, and a UWB module based on the IEEE 802.15.4-compatible Qorvo DWM1001C radio module \cite{Rub1} communicated to the main board of an AGV through a USB-UART interface module. UWB sensors operate in bands 3.7-4.2 GHz with a center frequency of 3.9 GHz and a bandwidth of 500 MHz. The AGV is also equipped with an IMU sensor integrated with an Arduino Nano 33 IoT kit \cite{IMU_sensor} for getting acceleration in 3D dimensions. We undertook an experimental investigation employing a UWB anchor configuration encircling a rectangular robotic laboratory space with approximated dimensions $11\times8.5\times3.3$ [m]. To evaluate the precision of the algorithm employed, we gathered ground truth data utilizing a Light Detection and Ranging (LiDAR) sensor affixed atop an AGV with a millimetre-level accurate level, which is positioned to match the coordinates of the UWB tag. In Fig.~\ref{fig_ENV}, a snapshot of the trajectory and the layout of the UWB anchors is presented.

\textbf{GNN training and testing}: We trained our GNN on collected data from experimental measurements. The datasets were randomly divided into training (80\%) and validation (20\%) data. 
For the GNN architecture, $M_c(\cdot)$ uses an MLP with $\{3,16,64\}$ neurons per layer. $M_u(\cdot)$ and $R(\cdot)$ uses MLP structures of $\{66,32,16\}$, $\{18,8,2\}$ respectively. We consider a batch size of $32$, step size, $\varepsilon = 0.001$ and error tolerance $\epsilon = 1 \times 10^{-6}$. We use an Adaptive Moment Estimation (ADAM) optimizer in training. Our implementation is based on a Pytorch-geometric Python machine learning framework. The parameters of the GNN were carefully selected based on numerical validation.


\begin{figure}[t!]
	\centering
	\includegraphics[trim=0cm 0cm 0cm 0cm, clip=true, width=0.445\textwidth]{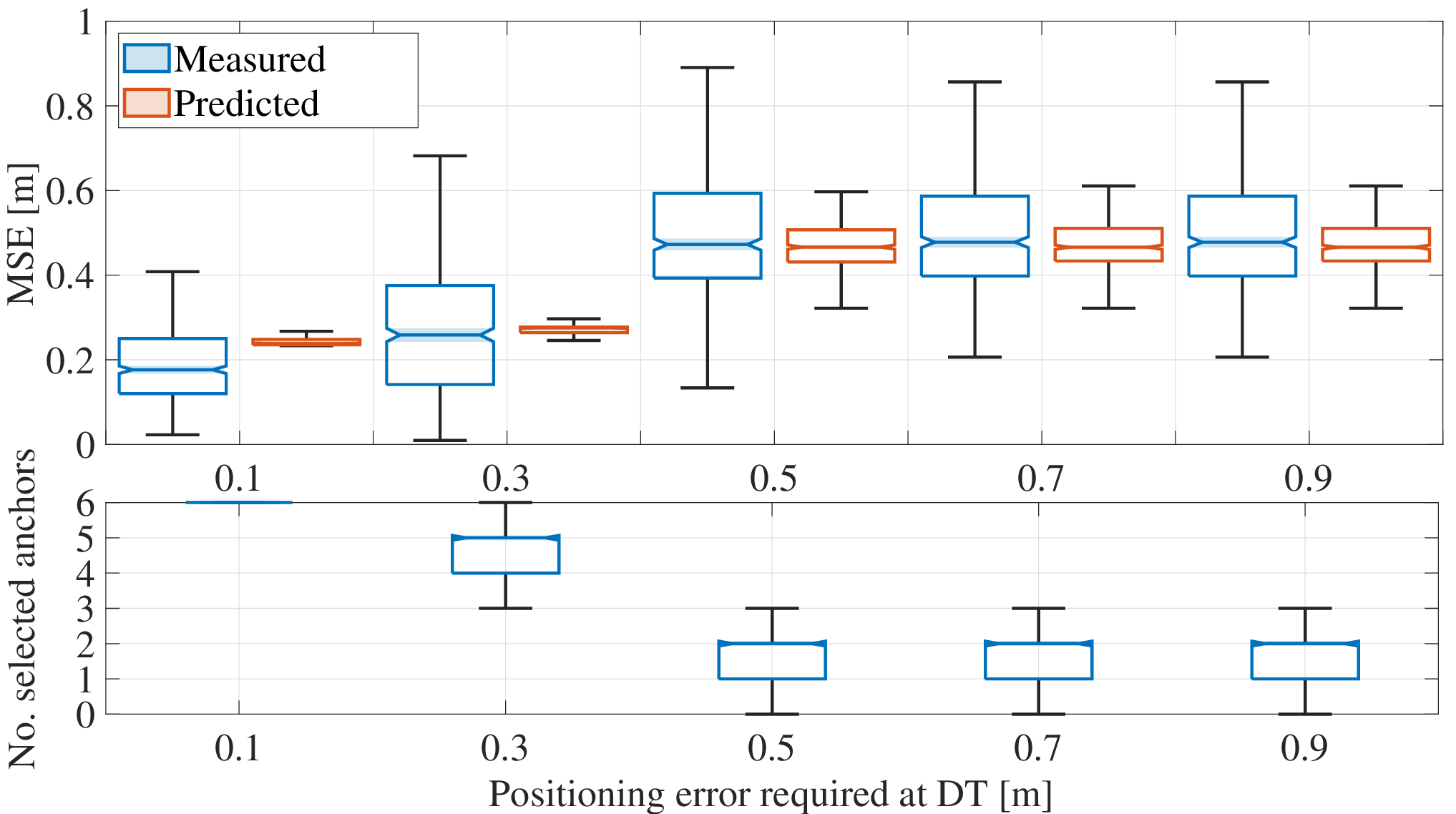}
        \vspace*{-0.3cm}
	\caption{The MSE and average number of selected anchors per QI vs positioning error requirement at DT.}
	\label{fig_MSE}
	\vspace*{-0.6cm}
\end{figure}
In Fig.~\ref{fig_trajectory}, we illustrate a setup wherein the AGV navigates along a rectangular path within an array of UWB anchors, delineating its trajectory with a blue line. Both Alg. 1 and GNN solutions exhibit proficient object-tracking capabilities in this setup. Notably, when receiving complete observations from all six anchors, the GNN model demonstrates exceptional accuracy, boasting an average error of approximately 5 cm. Alg. 1 not only provides precise location data but also offers estimations of error margins, along with recommending an optimal strategy for selecting UWB anchors. This multifaceted functionality is intuitively depicted in Fig.~\ref{fig_uncertainty_vs_QI}, wherein the evolution of uncertainty and the number of selected UWB anchors per QI are showcased across 300 QI test cases. These selections are based on their contributions to the performance of the DT. It is worth noting that the management of DT uncertainty is contingent upon DT requirements as outlined in \eqref{qos_condition}. Consequently, the DT only solicits additional observations when the predefined threshold is exceeded or when the AGV moves critical areas such as curves.

Fig.~\ref{fig_MSE} compares the 
measured Mean Squared Error (MSE) with the predicted MSE using Alg.~1 across various positioning error requirements. We also demonstrate the corresponding selection strategy for $\SA$s. Remarkably, the predicted error closely mirrors the actual error distribution compared to the measurement across all thresholds. When demanding a position accuracy of 10 cm, despite consistently selecting all six UWB anchors per QI, the DT accurately forecasts the error to be approximately 22 cm, a finding confirmed by the measured MSE, which stands at 19 cm. As the required accuracy threshold escalates, the number of selected $\SA$s gradually decreases, with our system adeptly forecasting the error while ensuring that the actual error distribution aligns with the requirements. Note that this is when an error greater than 50 cm is permissible; the DT framework necessitates only an average of 2 $\SA$s per QI while ensuring reliability. This emphasizes the reliability of our framework in real-world operating systems. 

\begin{figure}[t!]
	\centering
	\includegraphics[trim=0cm 0cm 0cm 0cm, clip=true, width=0.35\textwidth]{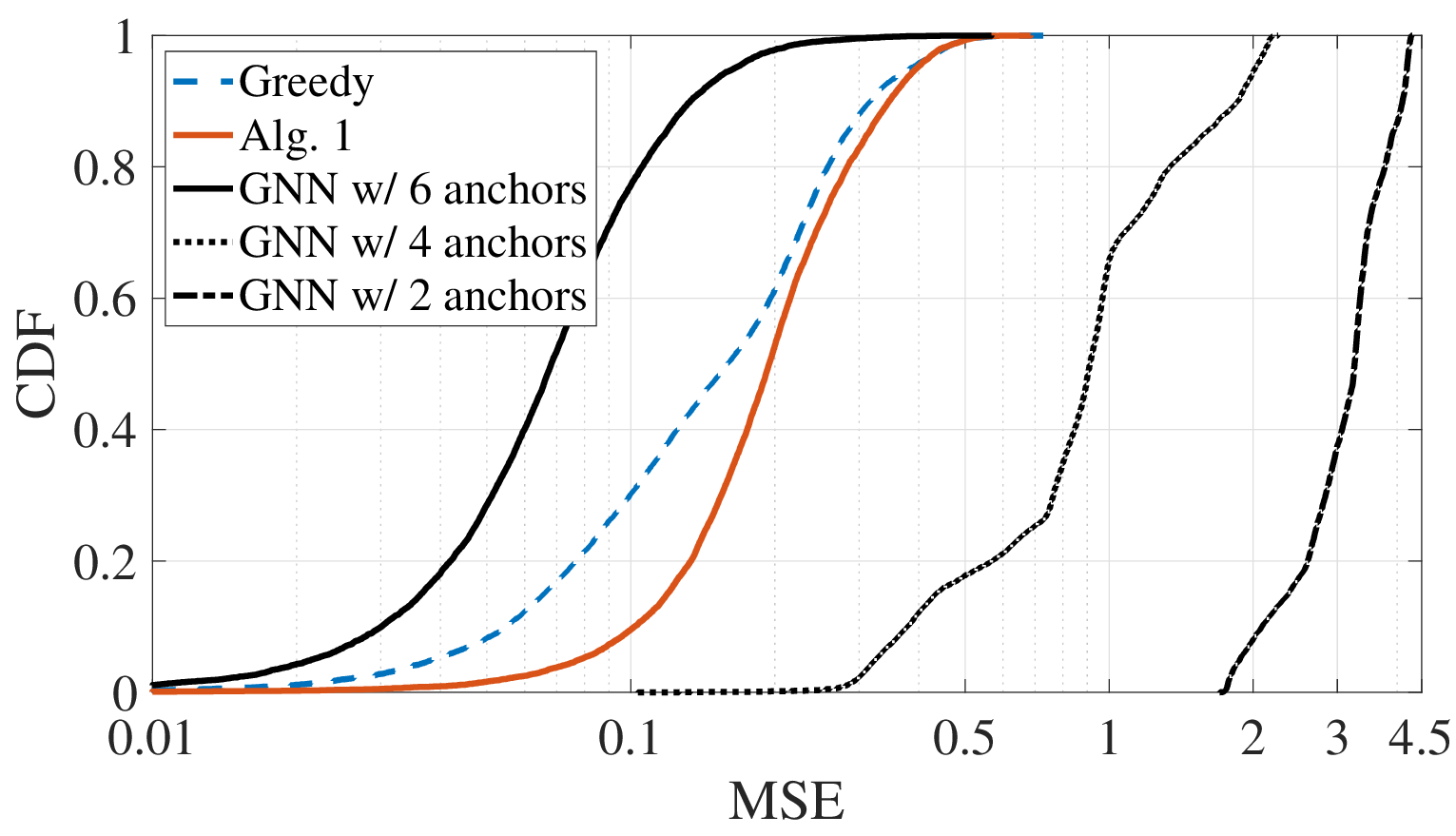}
        \vspace*{-0.25cm}
	\caption{The CDF of MSE across different schemes.}
	\label{fig_CDF}
	\vspace*{-0.6cm}
\end{figure}

Fig.~\ref{fig_CDF} compares the cumulative distribution function (CDF) in terms of measured MSE of Alg.~1 and the GNN approach against the \textit{Greedy} baseline, which enables the DT to access observations from all 6 $\SA$s at each QI. Despite Alg.~1 collecting fewer observations, it achieves a comparable average MSE to \textit{Greedy}, as 0.18 and 0.22, respectively. While the GNN method solely determines position without indicating uncertainty, it exhibits remarkable accuracy under favourable conditions, typically after gathering all 6 observations. This accuracy surpasses EKF methods, proving the superiority of machine learning-based approaches for addressing state monitoring challenges. However, its performance declines under more challenging conditions with reduced observations, particularly when observations contain substantial noise, and GNN is trained solely with the complete set of 6 anchors. This demands further research to enhance GNN's performance, aiming to improve accuracy with fewer observations through neural network architectures capable of effectively modelling time series data.

\vspace*{-5pt}
\section{Conclusion and Future Works}
We present a DT-based framework to optimize the UWB-based state monitoring, aiming to maintain the confidence of the system estimate. We also provided a low-cost DT-based positioning solution utilizing a GNN, ideal for scenarios that lacks a precise understanding of the system dynamics. We validated the results through real-world experiments conducted in the industrial laboratory. Our future work will extend the proposed framework to accommodate the dynamic nature of industrial environments, including moving and unknown obstacles.

\setstretch{0.9}
\bibliographystyle{IEEEtran}
\vspace*{-9pt}
\bibliography{Journal}
\vspace*{-12pt}

\end{document}